\documentclass{article}
\usepackage{graphicx} 
\usepackage{natbib}
\usepackage{array}
\usepackage{booktabs}
\usepackage{threeparttable}
\usepackage{multirow}
\usepackage{url}
\usepackage{lipsum}
\usepackage{amssymb}
\usepackage{amsmath}
\usepackage{mathrsfs}
\usepackage[hidelinks]{hyperref}
\usepackage{array}
\usepackage{booktabs}
\usepackage{threeparttable}
\usepackage{multirow}
\usepackage{verbatim}
\usepackage{xcolor}

\usepackage{geometry}
\newcommand{\virgolette}[1]{``#1''}
\newcommand{\review}[1]{{\color{black}#1}}

\title{Measuring frailty in older adults: an indicator based on a super-classifier \\[1ex] \large Preprint}
\author{
Sara Rebottini$^{1,2, \dagger}$,
Margherita Silan$^{3}$,
Pietro Belloni$^{3,}$\footnote{Correspondence to: pietro.belloni@unipd.it} $^,$\footnote{These authors contributed equally}
\\
\small $^{1}$ Department of Statistics, Computer Science, Application, University of Florence, Florence, Italy\\
\small $^{2}$ Faculty of Education, Free University of Bozen, Bozen, Italy\\
\small$^{3}$ Department of Statistical Sciences, University of Padova, Padova, Italy
}
\date{Version: October 2026}

\begin{document}

\maketitle

\section*{Abstract}
Identifying frail older adults in an ageing population is essential for improving healthcare services. This study proposes a composite indicator to assess individual frailty levels using administrative healthcare data. Given the complex and multidimensional nature of frailty, a multi-outcome approach is adopted. Following an extensive literature review, a set of adverse health events is selected as proxies for frailty. These events were modelled using logistic classifiers, with frailty determinants (associated to adverse health events, selected using a gradient tree boosting) serving as covariates. The sensitivity and specificity of each classifier is used to compose their combined likelihood. From this, we derive an indicator capable of quantifying frailty across the population. The indicator shows robust performance across multiple outcomes and over time. Its primary innovation lies in allowing the use of diverse and outcome-specific sets of frailty determinants without any structural constraint. Overall, we offer an effective tool for quantifying frailty among older adults, potentially supporting health authorities in the prevention of frailty-related adverse events.

\noindent
\textbf{Keywords}: Frailty, Classifier's combinations, Administrative data, Composite indicator

\maketitle
\section{Introduction}\label{sec.1}
The change in the age structure of the world's population has led to a widespread slow but steady ageing process, which translates into the need to adjust healthcare services provided to citizens.
Since the early 2000s, life expectancy at birth has increased from 66.8 years in 2000 to 73.1 years in 2019 \citep{WHO_world_2025}.
However, this increase was not matched by an equivalent rise in healthy life expectancy, which has grown from 58.1 to 63.5 years over the same period.
This gap between lifespan and healthspan (what has been termed the \textit{longevity paradox} \citep{garmany2021}) implies that an increasing share of older individuals spend a significant portion of their lives burdened by chronic conditions, disabilities and functional decline \citep{donno2024}.

The growing older adult population, increasingly exposed to chronic diseases and disability, puts considerable strain on healthcare systems worldwide.
Conditions such as cardiovascular disease, diabetes and neurodegenerative disorders become increasingly prevalent with age \citep{vos2017}, contributing to reduced independence, social isolation and lower quality of life \citep{vermeiren2016}.
In this context, the identification of the most vulnerable individuals becomes crucial for an effective healthcare planning and an efficient allocation of resources.

This vulnerability is encompassed by the concept of frailty.
Within the epidemiological literature, frailty is broadly understood as a multidimensional syndrome of decreased physiological reserve and increased susceptibility to adverse health outcomes, including hospitalisation, disability and mortality \citep{fried2001frailty, clegg_development_2016}.
Over the past three decades, several descriptive and operational definitions of frailty have been put forward. These definitions are typically grouped into three paradigms that differ according to their theoretical framework \citep{rockwood_what_2005}: the biomedical paradigm, the cumulative deficit paradigm and the bio-psycho-social paradigm.

According to the biomedical paradigm, frailty is defined as a biological syndrome characterised by decreased physiological reserve and reduced resistance to stressors, resulting from cumulative declines across multiple physiological systems, which in turn causes increased vulnerability to adverse health outcomes \citep{fried2001frailty}.
The physiological changes associated with ageing are central to this process and significantly increase the risk of falls, fractures, hospitalisation and mortality among frail older adults.
The most widely used measurement tool within this paradigm is the Fried Frailty Phenotype \citep{fried2001frailty}, which classifies an individual as frail when at least three of the following five criteria are observed: unintentional weight loss, exhaustion, muscle weakness, slowness and low physical activity. 

The cumulative deficit paradigm offers a broader perspective, defining frailty as \virgolette{a state of chaotic disorganisation of physiological systems that can be estimated by assessing functional status, diseases, physical and cognitive deficits, psychosocial risk factors, and geriatric syndromes, with the aim of building as complete a picture as possible of risk situations for adverse events} \citep{rockwood2007frailty}.
According to this approach, frailty occurs when deficits progressively outweigh positive resources, leading to functional decline and loss of autonomy.
The primary measurement tool associated with this paradigm is the Frailty Index (FI) \citep{mitnitski_accumulation_2001}, which is computed as the ratio between the number of deficits in an individual and the total number of deficits considered, typically ranging between 30 and 70 items spanning physical health, cognitive function, mental health, sensory impairments and social factors.
The FI provides a continuous score reflecting the overall burden of accumulated deficits, and has demonstrated robust predictive value for adverse outcomes such as mortality and hospitalisation \citep{rockwood2007frailty, chang2018relationship}.

The bio-psycho-social paradigm defines frailty as \virgolette{a dynamic state affecting an individual who experiences losses in one or more domains of human functioning -physical, psychological and social - which are caused by the influence of a range of variables and which increase the risk of adverse outcomes} \citep{gobbens2010search}.
This formulation explicitly acknowledges the multidimensional nature of frailty, integrating biological factors such as sarcopenia and hormonal imbalances with psychological components such as cognitive impairment, depression and anxiety, and with social determinants such as social isolation, financial hardship and lack of access to healthcare.
The Tilburg Frailty Indicator (TFI) \citep{gobbens_tilburg_2010} is the main measurement tool developed within this paradigm, assessing frailty across physical, psychological and social domains through a self-assessment questionnaire with 15 questions.

In this work we distinguish between the conceptual definition of frailty and its operational measurement. Conceptually, we define frailty as a dynamic state that increases the risk of adverse outcomes \citep{gobbens2010search}.
Operationally, since the frailty state is latent and cannot be directly observed in administrative data, we measure it through its consequences: the proposed indicator quantifies an individual’s exposure to frailty-related adverse outcomes.
The operational definition necessarily depends on the available data \citep{gobbens2010search}. Accordingly, the variables entering the indicator include both markers of the frailty state (diseases, disability, use of health services) and determinants that modify the risk of each outcome (age and sex).

Regardless of the paradigm adopted, the ability to identify frail individuals at the population level is important for the implementation of health policies.
In Italy, this need has been explicitly recognised in the National Plan for Chronicity \citep{Ministerodellasalute2022} and more recently in the National Recovery and Resilience Plan (PNRR, Decree No. 77, 2022), which call for the stratification of the population into homogeneous subgroups based on health needs, with the aim of designing targeted, personalised care pathways and optimizing the use of healthcare resources.
Population stratification, however, requires data that cover the assisted population as a whole, a condition that ad-hoc surveys (no matter how detailed on an individual basis) cannot fulfil by design.
Survey data allow for the application of comprehensive frailty measures across multiple theoretical paradigms and enable the investigation of causal relationships and risk factors; however, they are inherently limited in coverage and cannot support the systematic identification of frail individuals at the local health unit level.
Administrative healthcare data (routinely collected for administrative purposes such as hospital discharges, drug prescriptions, emergency room access and disease exemptions) offer complete population coverage, regular updating and accessibility to health system stakeholders \citep{silan2022construction}.
Their use, however, requires careful adaptation of frailty definitions to available proxies, as well as the development of robust case-identification algorithms \citep{canova2019systematic}.

The aim of this work is to construct a frailty measure for each individual assisted by the Local Health Unit serving the Province of Padova, in order to provide a practical support tool for prevention and public health planning.
The proposed measure is designed with the following key features:
\begin{itemize}
    \item Frail individuals are identified as those most exposed to adverse health outcomes associated with frailty (outcome-based definition); the measure is therefore also designed to be predictive of these events. The indicator is intended for population-level healthcare planning rather than individual clinical assessment.
    \item The indicator captures the multidimensionality of frailty while selecting the determinants that best predict the onset of each outcome using a different subgroup for each adverse event.
    \item Although the construction of the indicator simultaneously considers multiple adverse outcomes, the result is a single composite score, which can be easily used by healthcare practitioners to stratify the population and monitor frailty levels over time.  
    \item The indicator is built exclusively on administrative healthcare data, ensuring its ability to describe the frailty level of the assisted population as a whole without relying on ad-hoc surveys.
\end{itemize}

The article is organised as follows:
Section~\ref{sec.2} describes indicators used in the scientific literature;
Section~\ref{super-classifier} introduces a new approach for frailty indicators composition, focusing on the aggregation phase of the indicator construction;
Section~\ref{application} is the application of our methodology to real healthcare administrative data, including some preliminary descriptive consideration about the indicator;
Section~\ref{validation} describes and validates the indicator, showing its ability to predict the onset of adverse events;
Section~\ref{discussion_and_conclusion} provides several considerations about strength and limitations of our indicator proposal and possible future work.

\section{How to represent frailty}\label{sec.2}
\review{According to literature, frailty is} a multidimensional, complex and latent (not directly observable) concept.
These characteristics make it difficult to move from theoretical to practical definitions when it comes to frailty.
\review{As mentioned in the first section,} frailty can be defined as a health condition common in \review{older} individuals with an increased probability of observing adverse events that characterize this condition.
As argued in \cite{gobbens2010search}, the operational definition of frailty depends on the available data.
Administrative healthcare data represent a valuable resource for population-level health research, as they contain information on the entire assisted population at virtually no additional cost, since they are routinely collected for management and administrative purposes.
These data sources record a wide range of health-related events, including prescribed and dispensed medications, specialist outpatient visits, hospital admissions with associated diagnoses and more generally all accesses to healthcare services.
This breadth of information makes administrative data particularly suited for the construction of frailty indicators at the population level, as they allow to reconstruct the health profile of every individual assisted by a Local Health Unit through the record linkage of different data flows \citep{silan2022construction}.

However, since these data are collected for administrative rather than clinical purposes, their use in health research presents some challenges.
The coding systems adopted in administrative databases are designed to meet managerial and billing needs, and may not always capture the full clinical complexity of a patient's condition. As a result, the construction of valid proxies for health conditions (such as chronic diseases or functional limitations) requires the development of careful case-identification algorithms \citep{canova2019systematic}, which must account for the specific structure and limitations of each data flow.

Several indicators have been proposed in the literature to identify frailty using administrative data.
Among the most commonly cited indicators based on administrative data are the \textit{Hospital Frailty Risk Score} (HFRS) \citep{gilbert2018development}, the \textit{Elders Risk Assessment} (ERA) \citep{crane_use_2010}, the \textit{Frailty Index} (D-FI) \citep{drubbel_prediction_2013}, the \textit{Risk Prediction Model} (RPM-S) \citep{soong_dr_2019}, the \textit{Electronic Frailty Index} (eFI) \citep{clegg_development_2016}.
In addition, given its methodological and contextual relevance to the present work, we also consider the \textit{POSET Frailty Index} (FI-POS) \citep{silan2019quantifying, silan2022construction}, which shares the same conceptual framework and relies on the same type of administrative healthcare data flows used in this study.

The HFRS indicator aims to predict three outcomes: 30-day mortality, prolonged hospitalisation and emergency hospitalisation within 30 days.
Built using electronic hospital records of older individuals (aged 75 and above), the indicator is calculated as the sum of scores assigned to a person's diseases. 
These scores are proportional to each disease's ability to predict whether an individual belongs to a frailty cluster.
The clusters were identified through a cluster analysis, grouping individuals based on known frailty markers from the literature.
This indicator predicts adverse events with an Area Under the ROC Curve  (AUC) of 0.69 for 30-day mortality, 0.73 for prolonged hospitalisation, and 0.61 for emergency hospitalisation within 30 days.

The ERA is a prognostic indicator based on data from the Mayo Clinic (USA).
Constructed from information on demographic characteristics, health conditions and medical history relating to the two-year period 2003-2004, it aims to predict the identification of older adults at risk of hospitalisation or emergency room admissions in the following two years.
The main risk factors were identified using statistical models, each of which was assigned a score proportional to its estimated effect.
The frailty indicator varies between -7 and 32 and reflects the individual risk level.
However, the AUC of the model (0.678 overall, 0.640 for hospitalisation, 0.705 for emergency room admissions) indicates moderate predictive performance.

The D-FI is developed on a sample of 1679 Dutch subjects over the age of 60.
It aims to predict the occurrence of an adverse event between death, surgical emergency room visit, after-hours visit to the family doctor and admission to a nursing home.
The indicator values vary between 0 and 1 and are calculated as the number of deficits out of the total of 36 considered.
This indicator, with an AUC of 0.69, also shows moderate predictive ability of the risk of at least one adverse event occurring.

The RPM-S is constructed from clinical information in the UK Hospital Episode Statistics (HES) database, on a sample of 2099252 individuals aged 65 years and above.
This indicator is developed with the aim of predicting three adverse events: hospital mortality, emergency readmission within 30 days of discharge and institutionalisation.
To estimate the risk of these events, logistic regression models are used, which include certain a priori selected risk factors as predictors, such as gender, age, cognitive and functional problems, comorbidities and hospitalisation.
The indicator shows moderate predictive performance, with AUCs between 0.62 and 0.66 for hospital mortality, between 0.63 and 0.65 for institutionalisation, and between 0.57 and 0.63 for readmission within 30 days.

The eFI is constructed using medical record data from the databases of ResearchOne for index estimation and The Health Improvement Network for validation.
With the aim of predicting death, emergency admission and nursing home admission in several years' time, the selection of variables is based on the scientific literature and the deficit accumulation paradigm, excluding factors with low prevalence.
The final indicator includes 36 variables and assigns each subject a value between 0 and 1, calculated as the ratio between the number of deficits that are present and the total number of deficits.
The index shows good predictive ability over the years, variation in success over outcomes \citep{clegg_development_2016} and is now one of the most widely used in the literature to the point that it has been incorporated into guidelines for frailty management by the British Geriatrics Society and the National Health Service (NHS) \citep{turner_best_2014}.
Some extensions have been proposed in the United States \citep{pajewski_frailty_2019} and Japan \citep{nishimura_assessment_2022}.

The FI-POS aggregates 8 variables using the POSET method \cite{silan2025}.
The variables are selected through a series of logistic models from a large set of health-related features obtained by cross-referencing multiple healthcare databases.
A partially ordered set (POSET) refers to a set of elements where an order relation can be established. It is defined as \textit{partially} ordered because not all pairs of elements are comparable \citep{bruggemann2004estimation, bruggemann2011ranking, de2011approximation}.
Recently, the POSET method has been applied in various contexts, primarily in ranking theory, construction of composite indicators and causal inference in multi-treatment cases \citep{silan2021matching, silan2023identification}.
The FI-POS indicator has been evaluated in predicting the occurrence of six adverse events: death (AUC = 0.85), emergency room (ER) access with maximum priority (AUC = 0.81), hospitalisation (AUC = 0.66), femur fracture (AUC = 0.77), dementia onset (AUC = 0.81) and disability onset (AUC = 0.75).

All indicators described above, while contributing significantly to the literature, present some limitations that this work aims to address.
First, most of them are designed to predict a limited number of adverse outcomes, typically two or three, which may not be sufficient to capture the multidimensional nature of frailty \citep{gobbens2010search}.
For instance, the HFRS \citep{gilbert2018development} focuses on three outcomes related to hospitalisation, the ERA \citep{crane_use_2010} considers only hospitalisation and emergency room access, and the RPM-S \citep{soong_dr_2019} targets three post-hospitalisation outcomes.
A frailty indicator that simultaneously predicts a broader set of adverse outcomes, spanning death, functional decline, fractures and access to healthcare, may better represent the complexity of the frailty condition and support comprehensive public health planning.

Second, several of these indicators rely on a large number of variables (such as the D-FI \citep{drubbel_prediction_2013} and the eFI \citep{clegg_development_2016}, both based on 36 items).
Within the cumulative deficit paradigm, a large number of items is a requirement for validity: since the index is a proportion of deficits, its precision and stability depend on the number of deficits considered \citep{rockwood_what_2005}.
Our indicator pursues a different target: variables enter it through their contribution to the prediction of specific adverse outcomes rather than as deficits to be counted, so its validity is assessed in terms of predictive accuracy.
Within this framework, a parsimonious set of variables offers a practical advantage, since data availability may vary across Local Health Units and an indicator intended for routine use in public health settings should be easy to replicate.
This choice may involve a trade-off between practical usability and the breadth of the frailty dimensions captured.

Finally, the FI-POS \citep{silan2019quantifying, silan2022construction}, which represents the closest methodological reference to the present work, is constrained by a fundamental limitation of the POSET aggregation method: only ordinal variables can be included in the indicator construction, since the method requires all units to be sorted according to all variables.
This leads to the exclusion of potentially relevant variables, most notably sex, which is a well-established determinant of frailty \citep{fried2001frailty}.

The present work aims to overcome these limitations by proposing a frailty indicator built exclusively on administrative healthcare data that: (i) simultaneously predicts multiple adverse outcomes associated with frailty according to the literature, in order to better reflect its multidimensional nature; (ii) is constructed using a parsimonious set of variables, to ensure its replication across different Local Health Units; and (iii) is not restricted to ordinal variables, allowing to fully exploit administrative data information without excluding any variable a priori.

\section{Constructing a super-classifier for frailty representation}\label{super-classifier}
Key steps in constructing a composite indicator are selecting the variables to be \review{included} and the aggregation procedure.
To tackle both, we propose an innovative and efficient methodology using classification models as elementary indicators to predict the probability of occurrence of health adverse events.
Specifically, our main contribution lies in the aggregation phase, while for the selection step, we provide a preliminary proposal that can be customised depending on the context of the application.

\review{\subsection{Conceptual framework}\label{conceptual.framework}
Before introducing the technical details, we briefly outline the conceptual structure underlying the proposed framework.
Frailty is treated as a latent and multidimensional condition that manifests itself through multiple adverse health outcomes.
Rather than relying on a single outcome or on a simple aggregation of risks, the proposed approach models each outcome separately and then combines the resulting information into a single frailty indicator using a structured aggregation rule.
The following subsections describe how this idea is implemented in practice.

Each outcome captures a different aspect of frailty and therefore provides only a partial information about the underlying condition.
For this reason, the analysis follows two steps.
First, each adverse outcome is modelled separately using predictors that are relevant for that specific outcome: this allows different dimensions of frailty to be captured without imposing a common modelling rule across outcomes.
Second, the information produced by these outcome-specific models is aggregated into a single indicator, the aggregation is designed to reflect both the magnitude of the predicted risks and the reliability of each outcome as a marker of frailty.

This approach differs from simpler alternatives such as unweighted averages or indices based on a single outcome, which implicitly assume that all outcomes are equally informative and equally well measured.
By contrast, our aggregation explicitly accounts for heterogeneity across outcomes, creating an efficient measure of frailty which can be interpreted further.}

\subsection{Variable selection to reduce the number of frailty determinants}\label{variable.selection}
To reduce the number of frailty determinants used to predict the occurrence of an adverse event, we rely on a measure of variable importance generated by a gradient tree boosting.
\review{Since any statistical or machine learning model that provides a variable selection (or even variable ranking) could be used at this stage, our entire pipeline is not reliant on this specific model choice.
The purpose of the variable selection step is to provide a practical implementation, rather than constituting a defining component of the method.}

\review{We attempted alterative variable selections using lasso, random forests and gradient tree boosting (see Section \ref{determinants selection}). While lasso performs autonomous variable selection through its penalty term,  we implemented a binomial test for random forests and gradient tree boosting, originally developed for random forests by \cite{packageRandomExpl} and here extended to gradient boosting. We preferred to adopt gradient tree boosting for the main analysis since it yielded a more stable and parsimonious selection, while also offering computational efficiency and easily interpretable measures of variable importance used to rank frailty determinants for the final classification.

Boosting is an iterative procedure that combines multiple weak learners (typically shallow decision trees) to improve predictive performance}.
Each iteration is estimated based on a different data set: the first iteration starts with a model training on the original data set; in the following iterations, misclassified observations enter the training procedure with greater weight to improve the classification rule \citep{Friedman_greedy_2001}.
The final boosting classification rule is the combination of the models, trained at each iteration and weighted by a measure of misclassification: models with higher classification error will weigh less than those with lower classification error.
In particular, we use \texttt{XGBoost}, a specific training algorithm of the gradient tree boosting where, at each iteration, a second order Taylor's approximation is used to quickly optimise the loss function \citep{Chen_2016_Boostin}.
The choice of this algorithm is driven by its efficiency with sparse data.
\review{Variable importance is assessed through the frequency of selection of each predictor as a splitting variable across the ensemble. Intuitively, predictors that are frequently chosen for splits contribute more to reducing the loss function and are therefore more relevant for the classification.

The variable importance can be statistically tested to determine which variables are most significant, using an approach similar to that of \cite{rachid_zaim_binomialrf_2020} for random forests. 
Under the hypothesis of non-informativeness, each predictor has the same probability of being selected at a split which can be approximated as $1/P$, where P is the total number of predictors \citep{Ishwaran2010}.
Gradient boosting typically relies on weak learners: shallow trees with a limited number of splits. In our setting, we further restrict the model to \textit{decision stumps} (trees with a single split, described for example in \cite{iba1992induction,holte1993very}), ensuring that the probability of selecting a given predictor does not depend on previous splits.
Under this specification, a non-informative predictor does not provide a systematic reduction in the loss function across iterations and its selection can be treated as a random event.

Let $S_p$ denote the total number of times $X_p$ predictor is selected across all splits in the ensemble, and $N_{split}$ the total number of splits.
Under the null hypothesis of non-informativeness, $S_p\sim Bin(N_{split},1/P)$.
We can test the hypothesis that the probability associated with the $p$-th variable is the same as the one associated with the other variables $(H_0: \hat{\theta}_p = 1/P)$, whereas the alternative hypothesis states that this probability is greater than one of the other variables $(H_1: \hat{\theta}_p > 1/P)$. Further methodological details are provided in the supplementary material (Section S2).}

The binomial test uses the statistic 
\begin{align}
    \tag{1}
   \label{eq.1}
       \frac{\hat{\theta}_p-\theta_0}{\sqrt{\theta_0(1-\theta_0)/\review{N_{split}}}}
\end{align}
for testing the null hypothesis: under $H_0$, this statistic follows a normal distribution with mean $0$ and standard deviation $1$ \citep{conover1971practical}.
However, since the data approximately follow a binomial distribution, rather than using the normal approximation, we prefer to use the exact binomial distribution to test the hypothesis.
After performing the tests, we select the frailty determinants as those variables that are used most frequently as splits in the classification trees.

\subsection{Combining classifiers based on their predictive power for latent concepts}
Our goal is to construct a super-classifier capable of measuring a latent concept.
This concept can be measured by combining individual classifiers devoted to represent different aspects of the concept itself.
The super-classifier should, therefore, be able to classify observations with greater accuracy than an individual classifier, as it represents a combination of them and captures multiple aspects simultaneously.

Let $n$ be the sample size, $m$ the number of classifiers to be combined to obtain a super-classifier \review{and $I$ the unknown value of the latent concept.}
We define a classification rule as the algorithm that predicts the class of a new data using $k$ covariates, and the predicted label as the category ($-1$ or $1$) assigned to the new data by the classification rule. 
Following the work of \cite{parisi_ranking_2014}, we can reframe the problem into a label estimation procedure.
Given $x_i$, the $k$-dimensional vector containing the values of the covariates for the $i$-th subject, we express the observed label of each classifier as a function of this vector: $y_{ij}=f_{j}(x_{i})$, with $j=1,\dots,m$, and we can denote $\hat{y}_{ij}$ as the predicted label of the $j$-th classifier for the $i$-th subject.
\review{In this setting, $m$ denotes the number of adverse health outcomes and $k$ the number of frailty determinants.
Accordingly, $x_i$ represents the vector of frailty determinants for $i$ subjects. Since we estimate one outcome-specific classification rule $f_j(x_i)$ for each adverse health outcome, for each $j$ outcome, $y_{ij}$ denotes the observed adverse health outcome, while $\hat y_{ij}$ is its predicted value.}

Under two assumptions, we estimate the unknown value of $I$ ($-1$ or $1$) using the likelihood function.
The first assumption consists in the observations $x_{1,\dots,n}$ being independent and identically distributed realisations of the $p_X(x)$ marginal distribution.
The second assumption concerns the $m$ classifiers, that need to be conditionally independent \citep{dawid1979}.
\review{As in similar approaches developed in the \textit{learning from crowds} machine learning literature \citep{raykar2010learning}, this second assumption should be interpreted bearing in mind that the prediction errors of a given classifier should be independent of those of any other classifier, rather than as a strict structural property of the data-generating mechanism \citep{rodrigues2018deep, tanno2019learning, tanno2021uncertainty}.}

More formally, given the $i$-th data, for all $1 \leq u \neq j \leq m$ and for each of the two predicted labels $y_{iu},y_{ij} \in \{-1,1\}$, the following holds
    \begin{align}
    \tag{2}
        \label{eq.2}
        \Pr(\hat{y}_{iu} = y_{iu}, \hat{y}_{ij} = y_{ij} | I) = \Pr(\hat{y}_{iu} = y_{iu} | I) \times \Pr(\hat{y}_{ij} = y_{ij} | I),
    \end{align}
where $I$ represents the real value of the super-classifier, assuming it could be observed.

Therefore, the likelihood function is given by the product of the likelihood functions of each observation, so
\begin{align}
    \tag{3}
   \label{eq.3}
   \mathscr{L}(f_1(\textbf{x}), \ldots, f_m(\textbf{x}); I) = \prod_{i=1}^{n} \prod_{j=1}^{m} \Pr(f_j(x_i) | I_i).
\end{align}
We focus on each individual contribution to the overall likelihood, given by
\begin{align}
    \tag{4}
    \label{eq.4}
    \mathscr{L}(f_1(x_{i}), \ldots, f_m(x_{i}); \review{I_i}) = \prod_{j=1}^{m} \Pr(f_j(x_i) | I_i).
\end{align}

Both the super-classifier and the label of each classifier can assume only two values, $-1$ and $1$.
Therefore, we express \review{Equation (\ref{eq.4})} \review{analyzing the log-likelihood separately for possible values assumed by the super classifier}. \\

\review{For $I_i=1$, the log-likelihood becomes
\begin{align*}
   \label{eq.5}
   \tag{5}
   \ell_{+} &=\ell(f(x_{i1}), \ldots, f(x_{im}); I_i=1) \\ 
   &= \log\Big(\prod_{j=1}^{m} \Pr(f_j(x_i) | I_i=1) \Big) \\
   &=  \sum_{i|f_j(x_i)=1} \log\Pr(f_j(x_i)|I_i=1) + \sum_{i|f_j(x_i)=-1} \log\Pr(f_j(x_i)|I_i=1).
\end{align*}
Conversely, for $I_i=-1$
\begin{align*}
   \label{eq.6}
   \tag{6}
   \ell_{-} &=\ell(f(x_{i1}), \ldots, f(x_{im}); I_i=-1) \\ 
   &= \log\Big(\prod_{j=1}^{m} \Pr(f_j(x_i) | I_i=-1) \Big) \\
   &=  \sum_{i|f_j(x_i)=1} \log\Pr(f_j(x_i)|I_i=-1) +  \sum_{i|f_j(x_i)=-1} \log\Pr(f_j(x_i)|I_i=-1).
\end{align*}
}

We define the sensitivity and specificity of the \review{$j$-th} classifier in predicting the value of the super-classifier as
\begin{align}
\notag
    sens = \Pr(\review{f_j(x_i)=1}|y=1) \quad \text{and} \quad spec = \Pr(\review{f_j(x_i)=-1}|y=-1).
\end{align}
Therefore, \review{Equation (\ref{eq.5})} can be rewritten as
\review{
\begin{align*}
    \ell_{+} = \sum\limits_{i|f_j(x_i)=1} \log sens_i+\sum\limits_{i|f_j(x_i)=-1} \log (1-sens_i),
\end{align*}
and Equation (\ref{eq.6}) as 
\begin{align*}
    \ell_{-} = \sum\limits_{i|f_j(x_i)=1} \log (1-spec_i)+\sum\limits_{i|f_j(x_i)=-1} \log spec_i.
\end{align*}
}

If the value of the super-classifier is $1$, \review{$\ell_+$ will be greater than $\ell_-$}.
On the other hand, if the value is $-1$, \review{$\ell_-$ will be greater than $\ell_+$}.
Therefore, we estimate the value of the super-classifier by identifying which of the two \review{components}  has the higher value, and this corresponds to evaluating the sign of the difference between \review{$\ell_+$ and $\ell_-$}
\begin{align*}
\tag{7}
\label{eq.7}
I_i &= \operatorname{sign}\Big(\sum_{i|f_j(x_i)=1} \log sens_i+\sum_{i|f_j(x_i)=-1} \log (1-sens_i)-\sum_{i|f_j(x_i)=1} \log (1-spec_i) -\sum_{i|f_j(x_i)=-1} \log spec_i\Big) \\
&=\operatorname{sign}\Big(\sum_{i|f_j(x_i)=1} (\log sens_j - \log(1 - spec_i)) +\sum_{i|f_j(x_i)=-1} (\log(1 - sens_j) - \log spec_j)\Big).
\end{align*}
To simplify this expression, we represent both summations using the two indicator functions
\begin{align}
\tag{8}
\label{eq.8}
    \frac{1 + f_j(x_i)}{2} = 
\begin{cases} 
0 & \text{if } f_j(x_i) = -1 \\
1 & \text{if } f_j(x_i) = 1 
\end{cases}
\text{\space\space and \space \space}
    \frac{1 - f_j(x_i)}{2} = 
\begin{cases} 
1 & \text{if } f_j(x_i) = -1 \\
0 & \text{if } f_j(x_i) = 1 
\end{cases}.
\end{align}
So, from \review{Equation (\ref{eq.7})}, we obtain
\begin{align*}
\tag{9}
\label{eq.9}
    I_i &= \operatorname{sign}\Big(\sum_{i=1}^{m} \frac{1 + f_j(x_i)}{2}(\log sens_j - \log(1 - spec_i)) +\sum_{i=1}^{m}  \frac{1 - f_j(x_i)}{2}(\log(1 - sens_j) - \log spec_j)\Big) \\
    &=\operatorname{sign}\Big(\sum_{j=1}^{m} f_j(x_i)\log\alpha_j + \log\gamma_j\Big),
\end{align*}
where 
\begin{align}
\tag{10}
\label{eq.10}
\alpha_j=\frac{sens_j \ \cdot \ spec_j}{(1-sens_j)(1-spec_j)} \quad \text{and} \quad \gamma_j=\frac{sens_j(1-sens_j)}{spec_j(1-spec_j)}.
\end{align}

This method works if both the classifiers and the super-classifier are binary, with $-1$ or $1$ values.
However, we can describe a broader situation with both continuous classifiers and a continuous super-classifier.
Therefore, if we assume
\begin{align}
\notag
    f_j(x_i) \in [0;1] \quad \text{\space\space and \space \space} \quad
    I_i \in (-\infty; +\infty)
\end{align}
we need to rewrite $f_j(x_i)$ and $I_i$ as binary variables to trace back to the previous case.
We define two new functions
\begin{align}
\tag{11}
\label{eq.11}
    g_j(x_i) = 
\begin{cases} 
-1 & \text{if } 0 \leq f_j(x_i) < c_j \\
1 & \text{if } c_j \leq f_j(x_i) \leq 1 
\end{cases}
\text{\space\space and \space \space}
    \mathscr{I}_i= 
\begin{cases} 
-1 & \text{if } I < d \\
1 & \text{if } I \ge d
\end{cases}
\end{align}
\review{where $c_j$ and $d$ are constant thresholds.
$c_j$ is calculated from the data (see Section \ref{parameters calculation}), where it is chosen to optimise the predictive performance of each outcome-specific classifier.
$d$, instead, plays only an auxiliary role in linking the continuous and binary formulations and is not required in the final continuous specification of the super-classifier.}
Using these two functions, we use the previous calculations and obtain the super-classifier
\begin{align*}
\tag{12}
\label{eq.12}
     \mathscr{I}_i &= 
    \operatorname{sign}\Big(\sum_{j=1}^{m} g_j(x_i)\log\alpha_j + \log\gamma_j\Big).
\end{align*}
Since the only observed value is the probability predicted by the classifiers, we express $g_j(x_i)$ as a function of $f_j(x_i)$ using the sign function and the $c_j$ constants
\begin{align*}
\tag{13}
\label{eq.13}
    \mathscr{I}_i &= 
    \operatorname{sign}\Big(\sum_{j=1}^{m} \operatorname{sign}\big( f_j(x_i)-c_j \big)\log\alpha_j + \log\gamma_j\Big).
\end{align*}

Suppose that $f_j(x_i)$ is a monotonically non-decreasing function of the $k$ covariates.
Under this assumption, we ignore the sign operator (which only returns the positive or negative sign) and obtain continuous values for $f_j(x_i)-c_j$.
Similarly, we interpret $\sum_{j=1}^{m}(f_j(x_i)-c_j)\log(\alpha_j)+\log(\gamma_j)$ as a monotonically non-decreasing function of the probability that the sign function of each classifier $f_j$ is equal to one.
Therefore, we estimate the continuous value of the super-classifier ($I_i$) as follows
\begin{align*}
\tag{14}
\label{eq.14}
I_i = \sum_{j=1}^{m} \big (f_j(x_i)-c_j \big )\log\alpha_j + \log\gamma_j.
\end{align*}

where $\alpha_j$, $\gamma_j$ are derived from sensitivity and specificity of $f_j$ using an approach similar to cross-validation.
Dividing the data into $l$ folds, for each fold a threshold $c_{lj}$ is selected as the value among $z$ candidate thresholds between 0 and 1 that maximises the sensitivity and specificity of $f_j$.
The final $c_j$ is obtained as the average of the $l$ selected thresholds.
Using the resulting $c_j$ as the threshold to dichotomise $f_j$, sensitivity and specificity are then calculated for each of the $l$ folds.
Their respective averages are used to calculate $\alpha_j$ and $\gamma_j$.
Further practical details of the estimation procedure are provided in Section \ref{parameters calculation}.

\review{Since Equation~(\ref{eq.14}) combines information from the outcome-specific models into a single frailty score, its structure can be better explained by viewing each outcome as a \textit{piece of evidence} about an individual's frailty.
Each model produces a predicted probability $f_j(x_i)$ that an $i$ individual  will experience a $j$ adverse event.
This probability is compared to a $c_j$ outcome-specific reference value, which can be interpreted as a neutral level of risk for that outcome.
Values above $c_j$ indicate higher risks than expected, while values below $c_j$ indicate lower risks than expected.
Therefore, the difference $(f_j(x_i)-c_j)$ measures how much an individual deviates from the assigned risk level  to $j$ outcome.
Since not all outcomes are equally informative about frailty (some outcomes are more closely linked to frailty than others), each deviation is weighted by a $\log \alpha_j$ factor which increases when the corresponding model has higher sensitivity and specificity (see Equation~(\ref{eq.10})).
Outcomes that are better indicators of frailty therefore contribute more to the final score, while noisier outcomes contribute less.
As a result, the indicator increases when multiple outcomes show higher risks than expected and when these outcomes are well predicted by the data, and it remains low when risks are consistently below typical levels.}

Finally, we obtain a measure with values in the $(0;1)$ interval.
We apply a \textit{min-max} transformation to normalise the super-classifier
\begin{align*}
\tag{15}
\label{eq.15}
 I_i = \frac{I_i-\min(I)}{\max(I)-\min(I)}.
\end{align*}
In conclusion, we have constructed a super-classifier capable of measuring a latent concept through a linear combination of individual classifiers.
Assuming that each individual classifier moves in the same direction, where $-1$ is associated with the best scenario and $1$ with the worst, we apply the same interpretation to the final combination: the closer the value is to $1$, the more the observations are associated with negative outcomes.

\section{Construction and use of the frailty indicator} \label{application}
\begin{figure}[t]
    \centering
    \includegraphics[width=\textwidth]{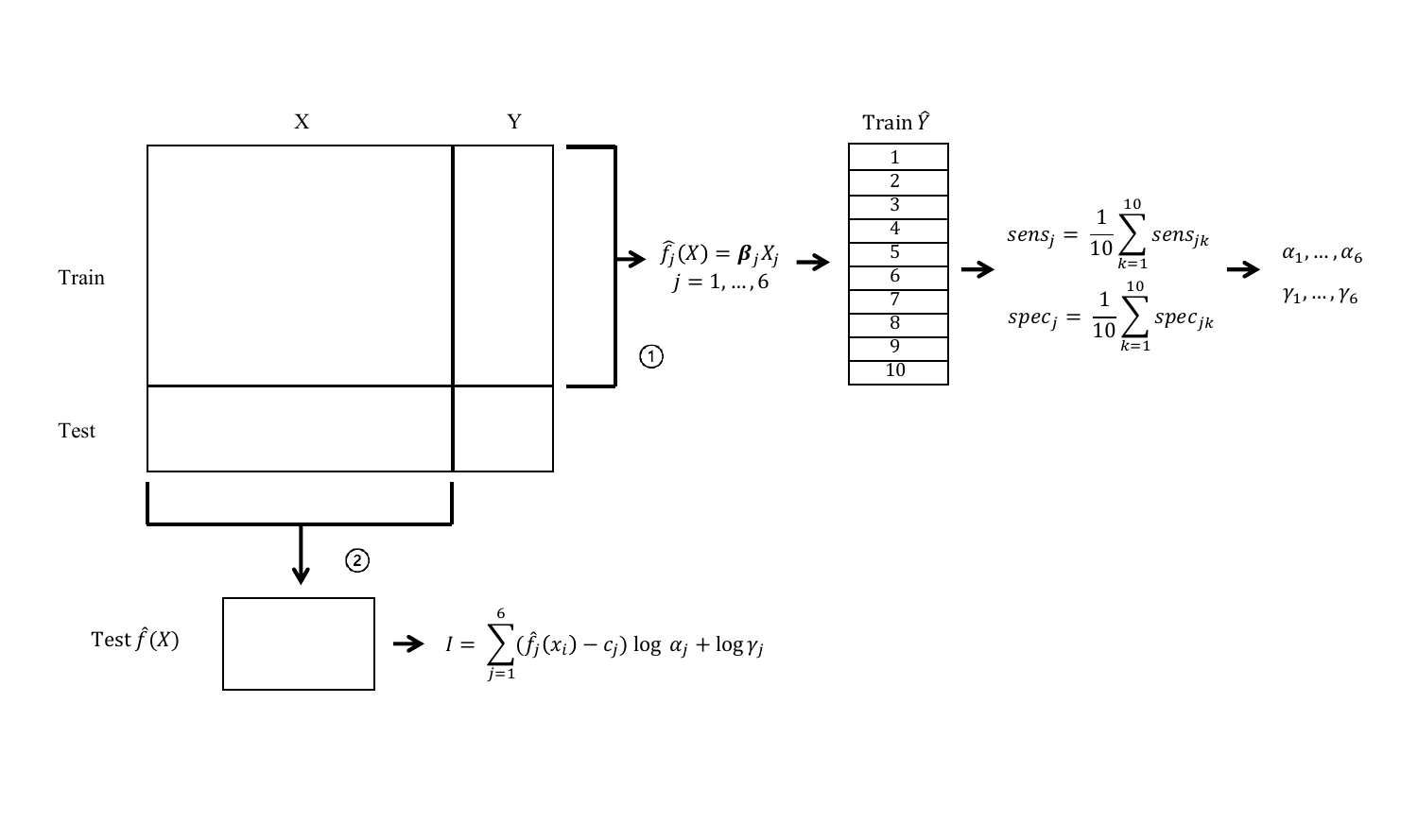}
     \caption{Data partition scheme for model estimation and parameter calculation. We use the training set to estimate the logistic regression models. For parameters calculation, we partition the training set into 10 groups. We use one group at a time as a test set, and, based on the previous models, we estimate the $\hat{y}_j$ occurrence probability of each outcome. We then calculate two accuracy measures for each group and we use the mean of the measures as sensitivity and specificity for each model. Finally, we combine these measures in order to calculate the $\alpha$ and $\gamma$ quantities (Step 1 in Figure). We use the test set for the calculation of the indicator: based on the values of the frailty determinants, the quantities previously obtained, and the outcomes' predicted probabilities, we compute the frailty indicator (Step 2 in Figure).}
    \label{fig:data_utilization}
\end{figure}

The construction of the indicator can be summarised in a step-by-step procedure:
\begin{enumerate}
\item The data sources are identified (Section \ref{data});
\item The selection of frailty determinants is carried out in four phases: a priori selection, exclusion of low-prevalence variables, exclusion of variables positively associated with the outcomes and application of the binomial test on the variable importance. See Section~\ref{determinants selection} for details;
\item The model is estimated by splitting the data into estimation and validation sets. Then, the $j$ desired outcome (with $j = 1, \dots, 6$) is identified and a logistic regression is fitted on the estimation set, which is balanced through under-sampling: 20\% of the sub-sample observations are equal to $1$ for $j$ outcome, and 80\% are equal to $-1$. This procedure is further discussed in Section \ref{Effect determinants};
\item The $\alpha_j$, $\gamma_j$, and $c_j$ quantities are calculated by partitioning the estimation set into ten sub-sets. For each unit in each group, the probability of developing the $j$ adverse event is computed. One $k$ group is chosen (with $k = 1, \dots, 10$) and the $c_{j_k}$ cut-off is selected as the value that maximises both sensitivity and specificity. The average of these cut-offs is then used to determine $c_j$. The predictive ability of each logistic regression model is evaluated using $c_j$ as the threshold for $j$ outcome. For each $k$ group of the estimation set, sensitivity, specificity, and AUC are computed, and average values of these parameters, reported in Table \ref{tab:parameters}, are used to determine $sens_j$ and $spec_j$. Finally, Equation (\ref{eq.10}) is applied to calculate the $\alpha_j$ and $\gamma_j$ quantities. This phase is illustrated in Section \ref{parameters calculation};
\item For each $j$ adverse event, the $f_j(x_i)$ probability of occurrence is computed for each individual. The $c_j$, $\hat{f_j}(x_i)$, $\alpha_j$, and $\gamma_j$ values are then plugged into Equation (\ref{eq.14}) and a single indicator score is obtained for each individual. The indicator score is then normalised using Equation (\ref{eq.15}), described in Section \ref{construction}.
\end{enumerate}
Given the complexity of the procedure for constructing the frailty indicator and the corresponding data use, we represent graphically in Figure \ref{fig:data_utilization} the data partitioning process.

\subsection{Data sources}\label{data}
\review{The development of a frailty indicator is motivated by a research collaboration with the Local Health Unit ULSS6 \textit{Euganea}, established through a formal agreement aimed at improving the identification of frail individuals within the assisted population.
This collaboration provides access to the administrative healthcare data flows of ULSS6 \textit{Euganea}, with the ultimate goal of supporting population stratification by frailty level and enabling the planning of targeted care pathways for vulnerable older adults.}
This institution oversees health services across the province of Padua, located in northeastern Italy, with the goal of providing public health services to the population.

Target population data come from the administrative health databases of the Epidemiological Unit of the Prevention Department.
We use data of two cohorts: \review{one consisting of individuals residing in the ULSS6 \textit{Euganea} catchment area from 1 January 2016 to 31 December 2017, and one from 1 January 2017 to 31 December 2018.}
The former consists of 213689 subjects and the latter of 216757 subjects who, as of 1 January 2018 and 2019, \review{are aged 65 or over.}
Frailty determinants were measured in 2016-2017 and 2017-2018, and adverse events in 2018 and 2019, respectively.
\review{For the sake of simplicity, the two cohorts are referred to as 2016–17→2018 (i.e. the determinants of frailty are measured in 2016–2017 and
the outcomes in 2018) and 2017–18→2019.

The fully anonymised dataset made available for this study draws from multiple administrative healthcare data sources:}
\begin{itemize}
    \item Regional health registry;
    \item Hospital discharge records;
    \item Emergency room (ER) admission records;
    \item Local psychiatry registry;
    \item Home care services database;
    \item Drug payment exemption database;
    \item Local pharmaceutical registry;
    \item Data from the local service for dependent older people.
\end{itemize}
From these databases, it is possible to derive information on 75 determinants of frailty and 9 adverse health events (outcomes).
We choose to use only 6 out of the possible outcomes: death, emergency room access with maximum priority, hip fracture, hospitalisation, onset of disability and onset of dementia.
We select this set of outcomes to remain consistent with other authors, who have extensively analysed the literature on the subject to choose which adverse events best represent frailty \citep{silan2025identifying}.
We use the second cohort \review{2017–18→2019} for the validation of the indicator and the first one \review{2016–17→2018} for the calculation of the quantities and the classification rule.
Therefore, we will refer only
to the second cohort of subjects in the validation phase.
\review{Many individuals are present in both cohorts, since they meet both inclusion criteria.
Consequently, to ensure the robustness of our findings against potential selection biases, we perform alternative validation strategies across different subsets of subjects. These approaches will be illustrated in Section~\ref{validation}.}

\subsection{Frailty determinants selection}\label{determinants selection}
We begin the \review{variables} selection process with 75 variables \review{cited in literature as risk factors of the considered outcomes} (see supplementary Table~S1).
These include demographic characteristics of individuals, binary variables indicating the occurrence of health conditions and count variables measuring the access to public health services.
First, we remove 9 variables due to their redundancy (duplicate variables, other composite frailty indicators, measures of health service duration). Since determinants with low prevalence do not help to explain the health of population as a whole, even if they might be relevant for a small group of individuals \citep{silan2022construction}, we exclude 19 determinants with low prevalence in our population (less than 1\%).
Subsequently, we discard the determinants with a positive association with each specific outcome, as we are interested in constructing a frailty indicator that detects only the deterioration of an individual's health status.
In order to identify the protective determinants, we take the odds ratio with each adverse event as a univariate measure of association, considering as protective factors those for which the odds ratio is significantly $<1$, with a 95\% confidence interval (see supplementary Table~S2).
For adverse outcomes such as dementia and disability, we conduct the analyses only in the sub-population in which the diagnosis of dementia and disability has not yet been made.
In order to better understand its effect on frailty, sex is always included in the selection regardless of its effect on the outcome.
Age and sex are not regarded as components of the frailty state, but as determinants that modify the risk of each outcome.
Age is not imposed a priori: it is retained by the selection procedure for every outcome and is among the strongest predictors (Table \ref{tab:logistic_models}).
Sex is retained by design because its effect differs not only in magnitude but also in direction across outcomes (Table \ref{tab:logistic_models}), so omitting it would distort the outcome-specific risk estimates on which the indicator is built.
After this initial selection, we obtain different groups of frailty determinants for each outcome.
Forty-three possible determinants are considered for death, 45 for hip fractures, 47 for access to the emergency room with maximum priority, hospitalisation and onset of disability and 44 for onset of dementia.

Subsequently, we proceed with the selection based on the binomial test.
As mentioned in Section~\ref{variable.selection}, \review{we applied three machine learning methods: lasso, random forests and gradient tree boosting.} To fit a model for each outcome, we choose a set of optimal hyper-parameters via cross-validation, reported in supplementary Table~S3. \review{ When implementing lasso, for each outcome the dataset was undersampled due to the significant outcome imbalance; otherwise, the outcome-specific models would have systematically reduced all coefficients to zero.
This adjustment was not required for random forests and gradient boosting, which have been shown to be more effective in handling unbalanced data.

We fix both the number of trees in the random forest and the gradient} boosting iterations at 2000 to allow low-prevalence determinants to be considered among the candidate split variables with a sufficiently high probability during the trees construction.
\review{While with lasso the subsample of determinants is obtained directly, for random forest and gradient tree boosting the importance measure of determinants is computed by multiplying the proportion of times it was used as a split by the total number of splits across all trees.}
Finally, we apply the binomial test with a 5\% significance level to identify the most important frailty determinants for each outcome.
\review{Using these methods (lasso, random forest and gradient tree boosting) we end up, respectively, with 36, 14 and 15 selected frailty determinants for death; 28, 16 and 16 for ER access with maximum priority; 37, 15 and 13 for hip fracture; 38, 20 and 16 for hospitalisation; 24, 14 and 12 for the onset of dementia; 40, 16 and 15 for the onset of disability.
Since gradient tree boosting was built to better tackle sparse and unbalanced data and provides a more parsimonious variable selection, we choose to rely on it for the frailty determinants selection.
More variable selection details are reported in supplementary Table~S4.}

\subsection{Effect of frailty determinants on adverse outcomes} \label{Effect determinants}
\review{Using the final sets of frailty determinants identified in Section~\ref{determinants selection}, we estimate the $f_j(\textbf{x})$ classification rules via logistic regression (one for each outcome), which serve as the building blocks of the frailty indicator.
The logistic models are estimated independently for each adverse event on 75\% of the population to avoid overfitting.
For disability and dementia, individuals who are already diagnosed are excluded prior to model estimation.
Given the low prevalence of several outcomes, each of the six models is trained on a data sub-set whose 20\% of the observations is positive to the adverse event and 80\% is not.

Although protective determinants were excluded a priori based on univariate associations (Section \ref{determinants selection}), additional issues may arise when variables are considered jointly in a multivariate setting.
In particular, some determinants may exhibit sign reversals or lose significance once included in the logistic models.
This behaviour does not reflect a genuine protective effect, rather arising from collinearity or interactions with other covariates.
This choice reflects a deliberate trade-off between predictive performance and interpretability: since the proposed frailty indicator is intended to capture a monotonic deterioration in health status, the presence of such unstable or sign-reversing effects would compromise both its interpretability and its internal coherence.

To address this issue, we conduct a diagnostic analysis aimed at identifying determinants whose multivariate behaviour is inconsistent with their univariate behaviour.
This step allows us to refine the set of covariates included in the final models.}
At first, for each outcome, we select the explanatory variables whose effects change in the multivariate context.
Then, we conduct a specific analysis for each of these variables.
We begin by training a model using only one of these variables as a covariate.
Based on the correlation matrix of the frailty determinants (see supplementary Tables from S5 to S10), we sequentially add the most correlated variables until the change in the effect is observed.
This way, we can understand which combinations of determinants, and in what order, modify the initial association between a determinant of frailty and the outcome.

Considering the death outcome, the first frailty determinant showing a reverse effect is related to endocrine disorders.
The sign reversal of the coefficient occurs when the variables related to diabetes and the number of multiple drugs prescriptions are included in the model.
This result is likely due to the structure of the variable associated with endocrine diseases, which is a binary indicator representing a macro condition: it takes the value 1 if at least one of several underlying conditions, including diabetes, is present.
As for multiple drugs prescriptions, the intake of a high number of medications is often a consequence of having multiple chronic conditions, thus explaining the correlation with endocrine diseases.
Then, we consider musculoskeletal diseases, which are significantly correlated with the number of emergency room visits.
The inclusion of this latter variable in the model is sufficient to cause a reversal in the effect on the risk of death.
In this case as well, the variable is aggregated and captures the presence of specific musculoskeletal conditions, including fracture diagnoses.
It is plausible that the observed effect reflects the tendency of individuals with suspected fractures to visit the emergency room for further evaluation.
As for urinary system diseases, no strong correlations (below 0.20) with other determinants emerges; however, a reversal in the effect occurs in the presence of a combination of variables: number of hospital admissions, emergency room visits, multiple drugs prescriptions, age and sex.

The same approach is also applied to the other adverse events, with the decision to analyse only the relationships that have not been investigated yet.
Regarding the outcome of emergency room visits with a maximum priority, we observe a reversal in the effect of digestive system diseases, particularly after the inclusion of variables such as number of hospital admissions, emergency room visits and multiple drugs prescriptions.
This reversal may reflect the frequency with which these conditions lead to acute episodes requiring urgent care, especially among older individuals \citep{annuario2020Online}.

A similar pattern is observed for hypercholesterolaemia: the inclusion of the aggregated variable related to endocrine diseases in the univariate model reverses its effect.
The cause may lie in the construction of the macro variable, which includes, among others, hypercholesterolaemia, diabetes and thyroid disorders, thus creating informational redundancy.
A reversal of effect is also found for the depression variable when the aggregated variable related to mental and behavioural disorders is included, as it also encompasses depression, making it difficult to isolate its specific effect.
Regarding the outcome of hip fracture, the inclusion of depression in the model reverses the effect of the aggregated variable for mental and behavioural disorders.

Furthermore, the number of hospital admissions shows a change in the effect when considered jointly with other relevant variables, such as age, previous fractures and depression.
In these cases, hospitalisation may represent a condition of monitoring and prevention, transforming the determinant from a risk factor into a potential protective factor.
With regard to hospitalisation, no similar sign reversals are observed.

For the onset of disability, a reversal in the effect of respiratory system diseases is observed when the multiple prescription variable is introduced. In this case as well, the explanation lies in the aggregated nature of the variable: the presence of multiple respiratory conditions may require numerous medications.

Finally, for the onset of dementia, an anomalous behaviour is identified for the number of poly-prescriptions and hospitalisations, which changes depending on the number of emergency room visits and the individual's age.

Based on the in-depth analysis on the variables with a protective effect, we exclude from the final model all determinants that show either a reversal of effect or a non-significant effect, even if protective in nature.
For categorical variables, the only retained determinants are the ones with at least one category with a statistically significant protective effect. 
The final estimated logistic models, along with the corresponding selection of frailty determinants, are presented in Table \ref{tab:logistic_models}.

\begin{table}[!t]
\centering
\tiny
\begin{tabular}{l|>{\centering\arraybackslash}p{1.1cm}>{\centering\arraybackslash}p{2.0cm}>{\centering\arraybackslash}p{1.5cm}>{\centering\arraybackslash}p{1.5cm}>{\centering\arraybackslash}p{1.5cm}>{\centering\arraybackslash}p{1.5cm}}
\hline
\textbf{Frailty determinants} & \textbf{Death} & \textbf{ER access with max. priority} & \textbf{Hip fracture} & \textbf{Hospitalization} &  \textbf{Disability onset} & \textbf{Dementia onset}  \\ 
\hline
Age {[}65-69{]} & ref.  & ref.  & ref.  &  ref. & ref. &   ref. \\
Age {[}70-74{]} & 1.53* & 1.36* & 2.65* & 1.08* & 1.19* & 2.83*  \\
Age {[}75-79{]} & 2.16* & 2.09* & 4.70* & 1.18* & 1.72* & 6.48*  \\
Age {[}80-84{]} & 4.03* & 3.06* & 9.24* & 1.31* & 3.33* & 13.63*  \\
Age {[}85-89{]} & 7.51* & 4.50* & 14.39* & 1.59* & 6.19* & 19.91*  \\
Age {[}90+{]} & 18.83* & 8.03* & 18.93* & 1.92* & 14.12* & 25.97*  \\
&  &  &  &  &  & \\
Atrial fibrillation &  &  &  & 1.16* & 1.23* &   \\
&  &  &  &  &  & \\
Cancer & 3.12* &  &  & 1.68* &  &  \\
  &  &  &  &  &  & \\
Depression &  &  & 1.66* &  &  &  \\
  &  &  &  &  &  & \\
Diabetes & 1.43* & 1.53* &  & 1.08* & 1.25* & 1.04*  \\
 &  &  &  &  &  & \\
Diseases of the digestive system &  &  &  & 1.04* &  &  \\
 &  &  &  &  &  & \\
Diseases of the nervous system & 1.66* & 1.60* &  & 1.14* & 2.25* &   \\
&  &  &  &  &  & \\
Fracture &  &  & 2.47* &  &  &  \\
  &  &  &  &  &  & \\
Sex [Female] & 0.67* & 0.69* & 1.72* & 0.78* & 1.07* & 0.90  \\
  &  &  &  &  &  & \\
Genitourinary diseases &  &  &  & 1.12* &  &  \\
&  &  &  &  &  & \\
Disability & 3.20* & 2.37* & 1.67* & 1.33*  &  & 2.14* \\
&  &  &  &  &  & \\
Mental and behavioural disorders & 1.32* & 1.32* &  & 1.09* & 1.62* & 2.75* \\
&  &  &  &  &  & \\
Multiple drugs prescriptions [0-3] &  &  &  & ref. & ref. &  \\
Multiple drugs prescriptions [4-6] &  &  &  & 1.29* & 1.22* &   \\
Multiple drugs prescriptions [7-9]&  &  &  & 1.48* & 1.37* &   \\
Multiple drugs prescriptions [10+] &  &  &  & 2.00* & 1.96* &   \\
&  &  &  &  &  & \\
N. of ER admissions [0] & ref. & ref. &  & ref. & ref. & ref. \\
N. of ER admissions [1-2] & 1.14* & 1.49* &  & 1.16* & 1.27*  & 1.29* \\
N. of ER admissions [3+] & 1.36* & 2.13* &  & 1.34*  & 1.28* & 1.65*  \\
&  &  &  &  &  & \\
N. of hospital admissions [0] & ref. & ref. &  & ref. & ref. &  \\
N. of hospital admissions [1-2] & 1.34* & 1.28* &  & 1.47* & 1.29 * &  \\
N. of hospital admissions [3+] & 2.36* & 1.51* &  & 2.49* & 2.30* &   \\
&  &  &  &  &  & \\
Respiratory system diseases & 1.47* & 1.96* &  &  &  &  \\
 \hline
\end{tabular}
\caption{Odds Ratio (with 5\% significance) of logistic regression models with the final selection of frailty determinants as covariates.}
    \label{tab:logistic_models}
\end{table}

Focusing on the coefficients of the estimated models, age emerges as the determinant with the strongest effect across all adverse events, while being female shows varying effects depending on the outcome.
In fact, sex acts as a protective factor for adverse events such as death, ER access with maximum priority and hospitalisation, but acts as a risk factor for hip fracture and the onset of disability.
No significant effect is observed, however, for the onset of dementia.
The frailty determinants associated with the outcomes of hip fracture, onset of disability and onset of dementia are the most affected by the variable selection process.
By the end of the selection procedure, only five variables for hip fracture, six for dementia onset and eight for disability onset are retained as sufficient to predict their occurrence.
This extensive reduction may compromise the indicator's ability to accurately detect the presence of these adverse events, with the risk of failing to identify some individuals as frail, even though they exhibit all the characteristics that would make them strong candidates.

\subsection{Parameters calculation} \label{parameters calculation}
Once the frailty determinants and the classification rules have been established, the next step consists in the calculation of $\alpha$, $\gamma$, and $c$ for the aggregation phase.
Data use is one of the main features of this phase, in order to prevent possible overfitting.
Given the low prevalence of some outcomes, it is not possible to split the data in two (one for the regression models estimation and another for calculating the parameters) with a classic train/test sets scheme, because there would be a robust probability that no event would be observed in either set.
Therefore, we use an approach similar to cross-validation.
We utilise the previously estimated models on the entire estimation set as classification rules, and we divide the estimation set into ten groups, treating each one in turn as a validation set.

To calculate $c_j$ for each outcome $j$, with $j=1,\dots,6$, we begin from a sequence of 500 values between 0 and 1. Then, we estimate the probability of developing the $j$ adverse event for the first subgroup of the estimation set.
Using each value of the sequence as a cut-off to dichotomise the probability of developing the $j$ adverse event, we choose $c_{j_1}$ as the cut-off for the first group, as the value that maximises the sensitivity and specificity of the classification rule.
We repeat the procedure for all ten groups and we use the average of $c_{j_1},\dots,c_{j_{10}}$ to obtain a final value for $c_j$.
For the calculation of $\alpha_j$ and $\gamma_j$, we proceed similarly: using $c_j$ as the threshold to dichotomise the probability of developing the $j$ adverse event, for each subgroup of the estimation set we calculate the sensitivity and specificity.
We then use the average of the 10 sensitivities and specificities to calculate $sens_j$ and $spec_j$ respectively, which we substitute into Equation (\ref{eq.10}) to determine $\alpha_j$ and $\gamma_j$.
The estimated parameters are reported in Table \ref{tab:parameters}.
We observe that the estimated models overall show good predictive capabilities, except for the ones employed for the prediction of hospitalisation and disability, which show the lowest values in terms of AUC.
This may be due to the selection of the determinants. Disability is among the outcomes with the fewest explanatory variables, while hospitalisation is a condition that may result from various factors not directly related to frailty.

\begin{table}[t!]
\centering
\begin{tabular}{l|c|ccc|cc}
\hline
\textbf{Outcome}& $c$ & \review{Sensitivity} & Specificity & AUC & $\alpha$ & $\gamma$ \\
\hline
Death & 0.196 &  0.793 & 0.800 & 0.868 &15.324 & 1.026  \\
ER access with max. priority & 0.196 & 0.755 & 0.765 & 0.823 & 10.032 & 1.029  \\
Hip fracture & 0.218 & 0.801 & 0.676 & 0.802 & 8.398 & 0.728  \\
Hospitalization  & 0.181  & 0.581 & 0.699 & 0.691 & 3.220 & 1.157 \\
Disability onset & 0.183 &  0.677 & 0.734 & 0.765 & 5.784 & 1.120  \\
Dementia onset & 0.218 & 0.789 & 0.736 & 0.834 & 10.425 & 0.857 \\
\hline
\end{tabular}
\caption{Values of the parameters calculated for the construction of the indicator.}
\label{tab:parameters}
\end{table}

\subsection{Super-classifier construction and normalisation} \label{construction}
\review{Before moving to the aggregation phase, it is necessary to check that the independence assumption between classifiers holds with regard to the estimated models.
In our context, this assumption deserves careful consideration, since all classifiers are applied to the same data and are estimated using partly overlapping sets of covariates derived from the same administrative sources.
In particular, the assumption would be violated if the models tended to misclassify systematically the same individuals (co-misclassification).
Nevertheless, some features of our modelling strategy mitigate this concern.
Each outcome-specific model is estimated on a different subsample of the population, constructed independently for each adverse event.
Since the outcomes are rare events and an undersampling procedure is adopted, only about 20\% of the observations are shared across the training sets of the different models.
Although the covariates are largely similar, estimation on partially distinct populations reduces the likelihood that common error structures arise from the estimation process.}

\review{Furthermore, we conduct an empirical assessment of potential dependence among the classifiers' errors.
Pearson residuals from each logistic model are computed on the external validation set.
We then examine the joint structure of these residuals across models using a hierarchical clustering procedure (results are reported in supplementary Figure S1).
No patterns of individuals misclassified by more than one model emerge.
While one model (the one related to hospitalisation) displays lower predictive performance, the corresponding misclassifications are not systematically shared with the other models.
This evidence suggests no indication of substantial co-misclassification that would undermine the conditional independence assumption.}

\begin{figure}[t]
    \centering
    \includegraphics[width=0.66\textwidth]{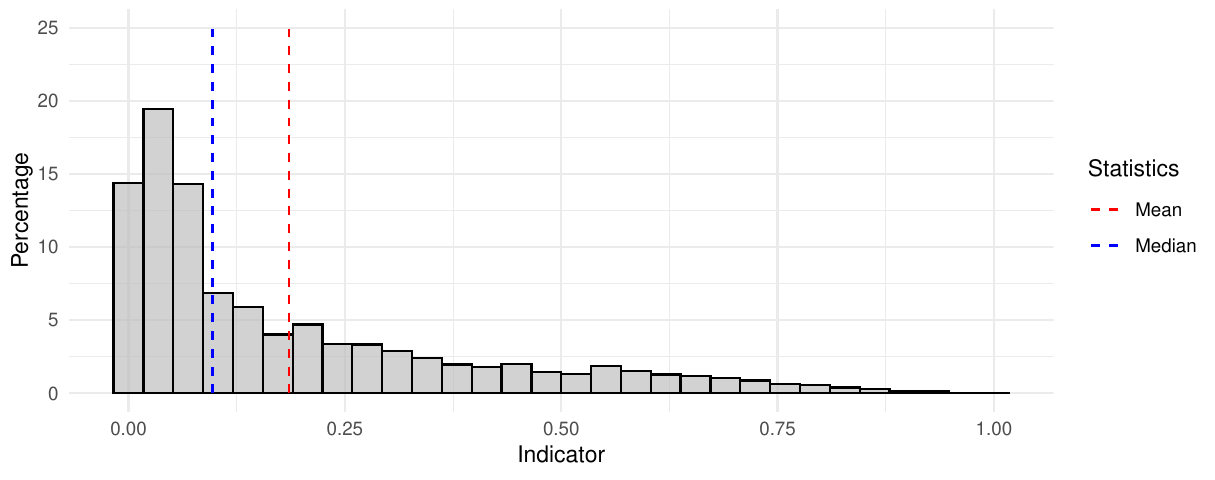}
     \caption{Distribution of the indicator calculated on the test set and some descriptive characteristics. About half of the population has a value of the indicator below 0.1 (median = 0.097). The mean value of the indicator in the whole population is 0.186 with a variance of 0.043, suggesting that only a small proportion of our population presents high values of the frailty indicator.
     }
    \label{fig:index_distribution}
\end{figure}

After calculating all quantities, we obtain the values of the indicator for each individual in the validation set.
We then use the outcome-specific models to estimate the probabilities that each individual will develop each outcome and, using the quantities reported in Table \ref{tab:parameters}, we estimate the frailty level for all individuals in the new set.
The distribution of the frailty indicator after the \textit{min-max} transformation is shown in Figure \ref{fig:index_distribution}.
The distribution highlights the presence of a large number of individuals with low frailty: approximately 50\% of the sample have an indicator smaller than 0.1.
These individuals are mostly under 75 years old and have not developed many of the health conditions included in the indicator yet.
Differences in the indicator based on gender are also observed: the average value of the indicator for women is 0.197, while for men it is 0.172 (for the entire population it is 0.186). Therefore, it appears that women have a greater frailty; this result is in line with other frailty indicators constructed with administrative data \citep{abrahaclaims2021, hall_development_2022, rockwood_what_2005}. However, this evidence might be due to differences in the age structure by sex \citep{silan2022construction}. To examine this, we standardise the frailty indicator by age and observe higher values for men than for women. Nevertheless, to remain consistent with the guidelines provided by \cite{rockwood_what_2005} regarding the characteristics a frailty indicator must have, we adopt the non-standardised version of the indicator in the subsequent analyses.

\section{Performance of the indicator and validation}\label{validation}
An essential part in the construction of a frailty indicator is verifying that it effectively captures the latent concept it was intended to measure.
Therefore, we base the validation of the proposed indicator on whether it exhibits the main characteristics that an efficient frailty indicator should show, as defined by \citet{rockwood_what_2005}.
In particular, we focus on assessing the \textit{criterion validity}, i.e., the indicator's ability to predict the adverse events related to frailty.
The indicator shows a high AUC in predicting five outcomes: death (AUC = 0.858), emergency room visits with maximum priority (AUC = 0.820), hip fracture (AUC = 0.785), the onset of disability (AUC = 0.760) and the onset of dementia (AUC = 0.818).
The only exception is hospitalisation, where the indicator shows a lower predictive ability (AUC = 0.655).
This can be justified by the multiplicity of causes that can lead to hospitalisation, such as planned surgery, trauma or acute infections.
In these terms, the proposed indicator seems to be consistent with other indicators in the literature based on administrative data \citep{silan2022construction}, and appears to correctly capture the concept of interest (Table \ref{tab:AUC_comparison}).

Despite the good predictive performance of the indicator, these results might be biased by overfitting the data.
Therefore, we re-assess the predictive ability of the indicator using data that had not been previously used. 
Specifically, we use the same data structure and population, but referred to a different time span.
This dataset contains determinants of frailty measured in 2017 and 2018, while adverse events are measured in 2019, out of 216757 observations.
Using the logistic regression models to estimate the probability of developing the adverse events for each new individual, and using the previously calculated $\alpha$ and $\gamma$ quantities, we compute the frailty indicator and evaluate its ability to predict the occurrence of each specific adverse event.
The observed AUC values are 0.857 for death, 0.776 for hip fracture, 0.828 for emergency room visits with maximum priority, 0.657 for hospitalisation, 0.807 for disability onset and 0.823 for dementia onset. These values confirm the indicator's good predictive performance even with a different data set.
In particular, we observe a slight improvement in predicting the onset of disability and dementia.

However, it is important to notice that this result may be influenced by the fact that the population did not change significantly from one year to the next \review{while their characteristics might have}.
Therefore, to further reduce the risk of overfitting, we repeat the analysis by selecting only those individuals who \review{were present in the 2017–18→2019 dataset and had also been part of the 2016–17→2018 validation set.}
This way, we ensured that the information used was not already incorporated in the previous indicator calculation.
Even in this case, the indicator shows good predictive ability: AUC is 0.856 for death, 0.834 for emergency room visits with maximum priority, 0.765 for hip fracture, 0.658 for hospitalisation, 0.802 for disability onset and 0.806 for dementia onset.

Finally, we use the indicator built with data from 2016 and 2017 to predict the onset of adverse events in 2019. \review{Despite a potential selection bias might arise from the death of some frail individuals, the AUC performance remains stable:} 0.832 for death, 0.816 for emergency visits with maximum priority, 0.750 for hip fracture, 0.642 for hospitalisation, 0.784 for disability and 0.796 for dementia.
We can thus conclude that the indicator maintains its predictive ability over time; given the way it is constructed, it would be possible to update $\alpha$ and $\gamma$ year by year using newly available information.
It would be interesting to assess the stability of the two quantities with other administrative data, preferably from different healthcare systems, to evaluate the indicator's dependence on the data used.

\begin{table}
\centering
\tiny
\begin{threeparttable}

\begin{tabular}{l|cc|ccc|cccccc}
\hline
\multirow{2}{*}{\textbf{Indicator}} & \multicolumn{2}{c|}{Data timespan} & \multicolumn{3}{c|}{Sample size} & \multirow{2}{*}{Death} & \multirow{2}{*}{\begin{tabular}[c]{@{}c@{}}ER access \\ with max. \\ priority \\ (AUC) \end{tabular}} & \multirow{2}{*}{\begin{tabular}[c]{@{}c@{}}Hip \\ fracture\end{tabular}} & \multirow{2}{*}{\begin{tabular}[c]{@{}c@{}} Hospi-\\talization\end{tabular}} & \multirow{2}{*}{\begin{tabular}[c]{@{}c@{}}Disability\\ onset\end{tabular}} & \multirow{2}{*}{\begin{tabular}[c]{@{}c@{}}Dementia\\ onset\end{tabular}}  \\ & \begin{tabular}[c]{@{}c@{}}Determ-\\inants\end{tabular} & \multicolumn{1}{c|}{\begin{tabular}[c]{@{}c@{}} Out-\\comes  \end{tabular}} & \begin{tabular}[c]{@{}c@{}} Disability\\ onset\end{tabular} & \begin{tabular}[c]{@{}c@{}} Dementia\\ onset\end{tabular}  &\begin{tabular}[c]{@{}c@{}} Other\\ adverse \\ events \end{tabular}& (AUC) & & (AUC) & (AUC) & (AUC) & (AUC)  \\ \hline
\review{Age-N} & \review{2016/17} & \review{2018} & \review{40865}  & \review{52291} & \review{53423} & \review{0.779} & \review{0.729} & \review{0.769} & \review{0.590} & \review{0.712} & \review{0.789} \\
FI-POS & \review{2016/17} & \review{2018} & \review{40865} & \review{52291} & \review{53423} & \review{0.854} & \review{0.805} & \review{0.765} & \review{0.664} & \review{0.749} & \review{0.805} \\
S-C & 2016/17  & 2018 & 40865 & 52291 & 53423 & 0.858 & 0.820 & 0.785 & 0.655 & 0.760 & 0.818  \\
S-C  & 2017/18 & 2019  & 166889 & 211877 & 216757 & 0.857 & 0.828 & 0.776 & 0.657 & 0.807 & 0.823  \\
\review{S-C} & \review{2017/18\tnote{$\dagger$}} & \review{2019} & \review{49604} & \review{61755} & \review{63020} & \review{0.869} & \review{0.843} & \review{0.791} & \review{0.667} & \review{0.816} & \review{0.833}  \\
S-C & 2017/18\tnote{$\dagger\dagger$} & 2019 & 39170 & 50049 & 51267 & 0.856 & 0.834 & 0.765 & 0.658 & 0.802 & 0.806  \\
S-C  & 2016/17\tnote{$\dagger\dagger$} & 2019 & 39170 & 50049 & 51267  & 0.832 & 0.816 & 0.750 & 0.642 & 0.784 & 0.796  \\ \hline
\end{tabular}
\begin{tablenotes}
\review{\item[$\dagger$] Subset limited to subjects appearing both in the 2016-2017 and the 2017-2018 test sets and new individuals not present in 2016-2017.}
\item[$\dagger\dagger$] Subset limited to subjects appearing both in the 2016-2017 and the 2017-2018 test sets. 
\end{tablenotes}
\caption{Comparison of the predictive capabilities, measured in area under the ROC curve, of the frailty indicator for the 6 adverse outcomes. The AUC for disability onset and dementia onset is calculated only among individuals without pre-existing disability or dementia, respectively. \review{Age-N: normalized age-class indicator (age-based proxy used as benchmark)}. FI-POS: \textit{POSET Frailty Index} \citep{silan2025}. S-C: indicator based on our super-classifier.}
\label{tab:AUC_comparison}
\end{threeparttable}
\end{table}

\begin{figure}[t!]
    \centering
    \includegraphics[width=0.6\textwidth]{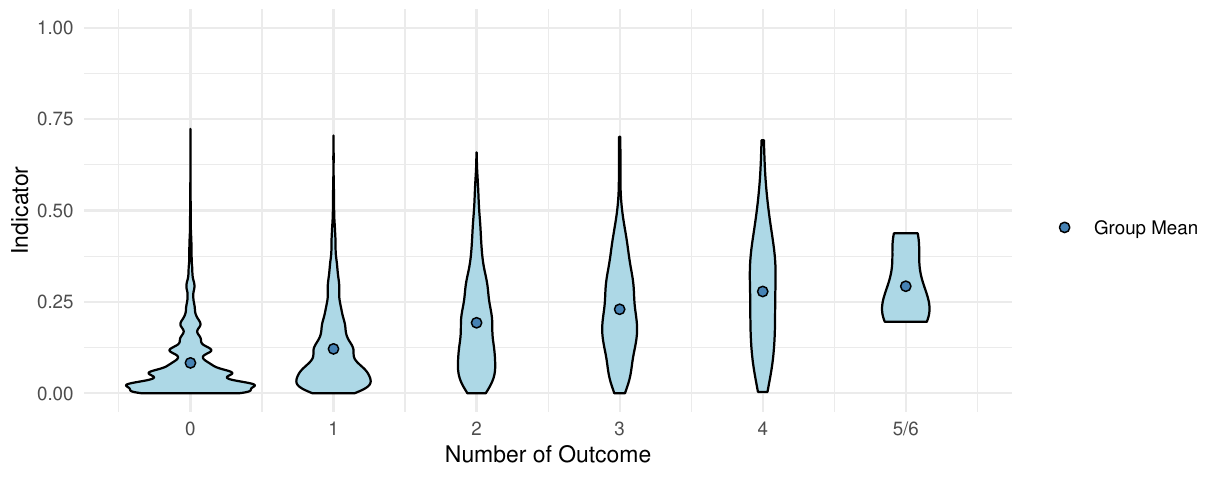}
     \caption{Distribution of the frailty indicator as the number of adverse events varies, using subjects from the 2018 test set who do not yet have a diagnosis of dementia and disability.}
    \label{fig:violin}
\end{figure}

We also need the frailty indicator to reflect the deterioration of older individuals' health.
Therefore, we expect that, as the number of adverse events increases, the value of the frailty indicator should also increase.
As shown in Figure \ref{fig:violin}, the frailty indicator appears to grow with the number of adverse health conditions, with its distribution showing higher values as the number of observed adverse events increases, as reflected in the rising average values.
When many outcomes are present, the frailty indicator does not increase linearly, but rather reaches a plateau.
This phenomenon, seemingly counter-intuitive, finds explanation in several well-established concepts in literature.
For example, the so-called selection effect, also known as survivor bias, whereby the frailest subjects often do not live long enough to accumulate a high number of outcomes \citep{aalen1994effects}.
Consequently, among those who end up having four, five or six outcomes, we generally find subjects who are certainly very frail, but who also present a certain physiological resilience and are able to survive long enough to accumulate more adverse outcomes.

So far, we identified a value between $0$ and $1$ that reflects the individual's frailty level.
However, we are not immediately able to classify a person as frail or not frail, and this is particularly relevant when identifying targets of interventions \citep{royston2006dichotomizing}.
To do this, it is useful to identify a threshold frailty level that allows for a distinction between frail and non-frail individuals.
Despite its importance, there is still debate in the literature about how to identify this threshold.
The aggregation method presented in Section \ref{super-classifier} allows the direct construction of a dichotomous indicator, if the formulation in Equation (\ref{eq.13}) is adopted instead of that in Equation (\ref{eq.15}).
Specifically, for each outcome, we assess the indicator's ability to correctly predict the occurrence ($1$) or non-occurrence ($-1$) of the adverse event using the F1 score (the harmonic mean of sensitivity and specificity) and the false negative rate (FNR) of classification (Table \ref{tab:ability}).

\begin{table}[t!]
   \centering
   \small
   \begin{tabular}{l|cccccc}
    \hline
      Score & Death & ER access with max. priority & Hip fracture & Hospitalization & Disability onset & Dementia onset\\
    \hline
     F1 & 0.200 & 0.072 & 0.029 & 0.337 & 0.232 & 0.072  \\
    FNR & 0.205 & 0.261 & 0.339 & 0.585 & 0.628 & 0.273 \\
    \hline
    \end{tabular}
    \caption{Performance of the binary indicator in predicting the adverse events occurrence: F1 score and the false negative rate of classification. The highest values of the FNR correspond to the outcomes with a low prevalence in the whole population.
    \label{tab:ability}}
\end{table}

Given the low false negative rates, it turns out that most of the classification errors are false positives.
Therefore, many individuals are classified as frail, and therefore at risk of adverse events, even though they do not actually experience them.
This suggests that our indicator tends to classify individuals as frail rather than as healthy.
However, in the context of frailty estimation, a high percentage of false positives is not excessively problematic, provided that the percentage of false negatives remains low.

Given the importance of maintaining a low number of false negatives, we finally assess whether the indicator systematically fails to identify specific individuals as frail, i.e., whether it excludes certain subgroups of the population more frequently.
Using the number of individuals classified as true positives, true negatives, false positives and false negatives stratified by frailty determinants and outcome-specific variables (see supplementary Table from S11 to S16), we evaluate whether the prevalence of each frailty determinant within the false negative group differs significantly from its prevalence in the general population.
This evaluation is based on Wilson confidence intervals, which offer a correction to classic Wald intervals and help address issues that may arise with small proportions \citep{wilson_interval_1998}.
We observe that, for the outcome of death, frailty determinants related to blood diseases and the hematopoietic system are more prevalent among false negatives.
In the case of emergency room visits, respiratory system diseases also show a higher prevalence in this group.
For hip fractures and the onset of disability, nervous system diseases appear more frequently among those the indicator fails to identify as frail.
Regarding the onset of dementia, mental and behavioural disorders are more common in the false negative group. Finally, for hospitalisations, individuals affected by neoplasms, cancer and mental and behavioural disorders are disproportionately represented among the false negatives.

\section{Discussion and conclusion}\label{discussion_and_conclusion}
\subsection{Key strengths}
In this work, we propose a new method to construct a frailty indicator using administrative data with the aim of stratifying the population according to the frailty status of each individual.
Our main contribution lies in the indicator construction pipeline, specifically in the frailty determinants aggregation phase.
The indicator is based on a combination of classification models (classifiers), each of which predicts the probability of developing a specific adverse event associated with frailty.
The accuracy measures of these classifiers, in particular sensitivity and specificity, are used as weights for the final aggregation.
Since the classification rules are independent, we can include all outcomes that we consider important in explaining the frailty condition (six, using the data described in Section \ref{data}).
In addition, we can use a small set of determinants, specific for each outcome, to predict the occurrence of the adverse event: 10 determinants are used to predict death, 9 to predict emergency room access, 5 to predict hip fracture, 13 to predict hospitalisation, 6 to predict the onset of dementia and 9 to predict the onset of disability.
This procedure allows us to differentiate and characterise the explanatory variables, generating a more accurate indicator.
For the frailty determinants selection process, we use a gradient tree boosting, but it is also possible to use any method capable of performing a variable selection.
The freedom in choosing the selection method and the number of outcomes to be considered allow this methodology to be improved and also to adjust to different needs.
In addition, selecting a different subgroup of frailty determinants for each outcome allows us to include variables whose effects on the probability of experiencing adverse events differ depending on the outcomes.
For instance, women are more likely than men to experience a hip fracture or the onset of dementia, while they are less likely to develop other adverse outcomes.

Comparing our indicator to the FI-POS indicator (a good benchmark among the indicators obtained from administrative data), our proposal is capable of including the sex of the individual, and potentially any other non-ordered variable, into the set of frailty determinants.
We also tried to construct the frailty indicator again excluding sex from the set of frailty determinants to better compare the AUCs with those obtained from the FI-POS indicator.
This is also useful to understand how important the contribution of sex is in our frailty estimation process. 
If we exclude sex, we obtain slightly lower values of AUC in all the outcomes (0.857 for death, 0.819 for emergency room visits with maximum priority, 0.654 for hospitalisation, 0.761 for disability onset and 0.817 for dementia onset), with the exception of the hip fracture (AUC = 0.787) adverse event.

\subsection{Limits and improvements}
As already mentioned, our indicator's prediction of the hospitalisation event shows a lower AUC compared to the other events' prediction, suggesting that the model may not be optimal for predicting this outcome, as it is very generic and can have many different causes, not necessarily related to frailty.
Indeed, among the main causes of hospitalisation is trauma, a condition that also involves other spheres of the older person's health beyond frailty \citep{annuario2020Online}.
To improve the hospitalisation predictive capacity, it might be useful to include additional explanatory variables in the classifier; in particular, determinants variables not directly related to the frailty condition could be included, if useful to explain the process underlying hospitalisation.
However, it is worth noting that this limitation ultimately does not lie in the aggregation method but in the choice of outcomes.

At the variable selection stage, some problems could arise in using the binomial test to identify the most important determinants of frailty.
The test performed on the number of times a variable is used as a split in gradient tree boosting does not take into account the decrease in the variance of the residuals.
This could result in the selection of a determinant that is used many times as a split in the final part of the trees, but whose contribution to decreasing variance is less than a determinant used a smaller number of times as a split in the top part of the trees.
In addition, our final selection identifies a very small number of determinants for the hip fracture and onset of dementia adverse events, a number that may not fully describe the underlying frailty mechanism.
This over-selection could be due to the low prevalence of such outcomes in the population.
Therefore, a different methodology could be used to select the most important determinants of frailty that can better handle the low population prevalence of some adverse events.

Since age and sex are included among the frailty determinants, part of the indicator reflects non-modifiable demographic risk rather than the frailty state in a strict sense.
This choice is driven by the intended use of the indicator: health authorities need to rank the entire population aged 65 and over with a single score, and stratifying a posteriori by age or sex would produce a set of mutually incomparable sub-indicators.

Our work applied the frailty indicator exclusively to administrative data from a specific health authority.
To further evaluate its robustness, it would be better to test the indicator using different data.
Applying our pipeline to build the indicator to different administrative data should be straightforward, as the reduced set of variables we selected are commonly collected across health authorities.
This process may present additional challenges, but could provide further insights on the indicator's versatility and reliability.
Finally, in order to have greater consistency of the frailty indicator over time, it would be recommended to update $\alpha$ and $\gamma$ using new data, including consecutive data batches in the estimation set, then updating the indicator with the new information.
This could increase the predictive capabilities of the indicator, but could also increase its dependence on the data used.

\subsection{Conclusion}
With the increase in the older population worldwide, especially in Italy, there has also been an increase in the prevalence of age-related diseases.
These diseases reduce healthy life expectancy with important consequences on the use of healthcare and welfare resources.
Frailty is a condition typical of older individuals that involves multiple aspects of a person's life, starting with a deterioration in overall health, such as loss of autonomy, or increased vulnerability to the development of adverse events, such as hospitalisation, onset of dementia or disability.
Among the older population, frail individuals are those most in need of care, so it is essential to identify them early to implement prevention policies to improve their quality of life \citep{bergman_global_2002}.
Therefore, the use of a tool capable of measuring the level of frailty of individuals using the healthcare administrative data plays a fundamental role.

To this end, we develop an indicator to measure the frailty level in the older population based on administrative data.
Consistently with the definition adopted in Section 1, frailty is measured operationally through the individual’s exposure to frailty-related adverse events, which reflects the underlying dynamic state of the individual.
The indicator is based on the risk of adverse events like death, hip fracture, emergency room visits, hospitalisation, dementia and disability.
The frailty determinants are selected using the importance measure of a gradient boosting model, and we use a novel aggregation method to combine single-outcome classifiers, weighted by accuracy measures, to generate a frailty score between~0 and~1.
The indicator shows good predictive ability for adverse events in 2018 and 2019, comparable to other state-of-the-art indicators, such as the FI-POS.
It also shows the possibility to predict adverse events two years after the frailty determinant recording, confirming its robustness over time.
One of the main advantages of our method is the possibility to use non-ordered variables (such as sex) as frailty determinants, which was not possible with the FI-POS aggregation technique.
To conclude, our indicator can be an important improvement for healthcare authorities, as it helps to identify vulnerable individuals using common variables easily available from administrative data collected for other purposes, without increasing costs and resources.

\section*{Acknowledgments}
The authors are thankful to G. Boccuzzo (University of Padua) and M. Stival (Ca' Foscari University of Venice) for the valuable suggestions after reading the first draft of the manuscript, and to M. Nicolaio (University of Padua) for the intensive support on the databases used in this work.
The authors extend their gratitude to B. Bezzon (University of Padua) for linguistic support.

\section*{Funding}
P.B. and M.S. acknowledge funding from Next Generation EU [DM 1557 11.10.2022], in the context of the National Recovery and Resilience Plan, Investment PE8 – Project Age-It: \textit{Ageing Well in an Ageing Society}.
The views and opinions expressed are only those of the authors and do not necessarily reflect those of the European Union or the European Commission.
Neither the European Union nor the European Commission can be held responsible for them.

\section*{Software}
The R code in the repository \url{https://github.com/pietrobelloni/superFI} can be used to reproduce the results presented in the article.
For privacy reasons, sharing the original raw data is not possible. Therefore, in the repository we only provide the values predicted by the logistic models mentioned in the article, which we aggregate to obtain the final frailty indicator.
The variable selection and model estimation steps are thus not included in the code.

\clearpage

\bibliographystyle{abbrvnat}
\bibliography{reference}

\end{document}


\maketitle

\section*{S1. Supplementary tables and figures}

\begin{longtable}{c|c|c}
     \caption{ List of factors determining frailty before the selection process and their data prevalence. Prevalence is not calculated for multi-class variables.
    } \label{tabApp:Esplicative} \\

    \textbf{ID} & \textbf{Variable} & \textbf{Prevalence (\%)} \\
    \hline
    \endfirsthead
    \multicolumn{3}{l}%
    {\small{\begin{tabular}[l]{@{}l@{}}{\tablename\ \thetable{}}: (continued) List of factors determining frailty before the selection process and their data prevalence.\\ \hspace{4.8em}Prevalence is not calculated for multi-class variables.\end{tabular}
    }} \\
   
    \textbf{ID} & \textbf{Variable} & \textbf{Prevalence (\%)}\\
\hline
    \endhead
    \multicolumn{3}{r}{{\textbf{Continued on next page}}}
    \endfoot

    \endlastfoot
    1 & Gastroenteritis & 0.313 \\
    2 & Alzheimer & 0.321 \\  
    3 & Angina & 0.432 \\ 
    4 & Osteopathies & 0.460 \\ 
    5 & Viral hepatitis & 0.777 \\ 
    6 & Myocardial infarction & 0.813 \\
    7 & Femoral neck fracture & 0.904  \\
    8 & Bronchiectasis & 0.049 \\
\end{longtable}

\landscape
\scriptsize
\begin{longtable}{M{2.6cm}|ccc|ccc|ccc|ccc|ccc|ccc}
\caption{Univariate association between frailty determinants and outcomes reported as odds ratio (OR) and lower/upper bounds of its confidence interval.}
\label{tab: Odds}
\\
\textbf{Variables} & \multicolumn{3}{c|}{\textbf{Death}} & \multicolumn{3}{c|}{\textbf{ER access with max priority}} & \multicolumn{3}{c|}{\textbf{Hip fracture}} & \multicolumn{3}{c|}{\textbf{Hospitalization}} & \multicolumn{3}{c|}{\textbf{Disability onset}} & \multicolumn{3}{c}{\textbf{Dementia onset}} \\
 & \textbf{OR} & \textbf{Lower} & \textbf{Upper} & \textbf{OR} & \textbf{Lower} & \textbf{Upper} & \textbf{OR} & \textbf{Lower} & \textbf{Upper} & \textbf{OR} & \textbf{Lower} & \textbf{Upper} & \textbf{OR} & \textbf{Lower} & \textbf{Upper} & \textbf{OR} & \textbf{Lower} & \textbf{Upper} \\
\hline
\endfirsthead
\multicolumn{19}{l}{{\small{\tablename\ \thetable{}}: (continued) Univariate association between frailty determinants and outcomes reported as odds ratio (OR) and lower/upper bounds of its confidence interval.}}\\

\textbf{Variables} & \multicolumn{3}{c|}{\textbf{Death}} & \multicolumn{3}{c|}{\textbf{ER access with max priority}} & \multicolumn{3}{c|}{\textbf{Hip Fracture}} & \multicolumn{3}{c|}{\textbf{Hospitalization}} & \multicolumn{3}{c|}{\textbf{Disability onset}} & \multicolumn{3}{c}{\textbf{Dementia onset}} \\
 & \textbf{OR} & \textbf{Lower} & \textbf{Upper} & \textbf{OR} & \textbf{Lower} & \textbf{Upper} & \textbf{OR} & \textbf{Lower} & \textbf{Upper} & \textbf{OR} & \textbf{Lower} & \textbf{Upper} & \textbf{OR} & \textbf{Lower} & \textbf{Upper} & \textbf{OR} & \textbf{Lower} & \textbf{Upper} \\
\hline
\endhead
 \multicolumn{19}{r}{\textbf{Continued on next page}} \\
\endfoot

\endlastfoot
Age (classes) & 30.09 & 27.04 & 33.58 & 15.13 & 12.89 & 17.85 & 27.35 & 20.18 & 38.01 & 2.82 & 2.69 & 2.95 & 14.68 & 13.33 & 16.17 & 47.59 & 36.64 & 63.11 \\
Alzheimer & 7.79 & 6.51 & 9.34 & 4.52 & 3.25 & 6.28 & 2.55 & 1.36 & 4.77 & 2.10 & 1.78 & 2.47 & 11.48 & 8.14 & 16.20 & 12.95 & 9.33 & 17.97 \\
N° ER accesses with max priority & 8.31 & 7.52 & 9.16 & 11.15 & 9.76 & 12.68 & 2.14 & 1.44 & 3.03 & 3.80 & 3.50 & 4.12 & 3.46 & 2.80 & 4.22 & 4.15 & 3.36 & 5.06 \\
Disability & 8.96 & 8.52 & 9.42 & 6.62 & 6.11 & 7.17 & 3.53 & 3.16 & 3.95 & 2.30 & 2.25 & 2.36 & /// & /// & /// & 5.25 & 4.85 & 5.69 \\
Bronchitis & 6.52 & 5.88 & 7.23 & 10.06 & 8.80 & 11.50 & 2.30 & 1.62 & 3.28 & 3.95 & 3.65 & 4.28 & 4.08 & 3.39 & 4.92 & 2.82 & 2.24 & 3.57 \\
Dementia & 5.89 & 5.44 & 6.39 & 5.16 & 4.54 & 5.87 & 2.97 & 2.35 & 3.75 & 1.97 & 1.84 & 2.10 & 8.71 & 7.72 & 9.83 & /// & /// & /// \\
Heart failure & 7.34 & 6.85 & 7.86 & 7.25 & 6.53 & 8.06 & 2.76 & 2.22 & 3.44 & 4.10 & 3.89 & 4.32 & 4.51 & 4.02 & 5.06 & 2.74 & 2.34 & 3.21 \\
Bronchiectasis & 5.44 & 3.28 & 9.04 & 7.12 & 3.59 & 14.11 & 1.64 & 0.23 & 11.75 & 4.02 & 2.74 & 5.90 & 2.81 & 1.20 & 6.61 & 3.47 & 1.27 & 9.44 \\
\end{longtable}
\endlandscape


\begin{table}
\scriptsize
    \caption{Grid values and optimal outcome parameters (hyper-parameters) to of the gradient tree boosting.}
    \label{tab:xgb_param}
%
\end{landscape}

\clearpage

\begin{figure}[h]
    \centering
    \includegraphics[width=0.75\textwidth]{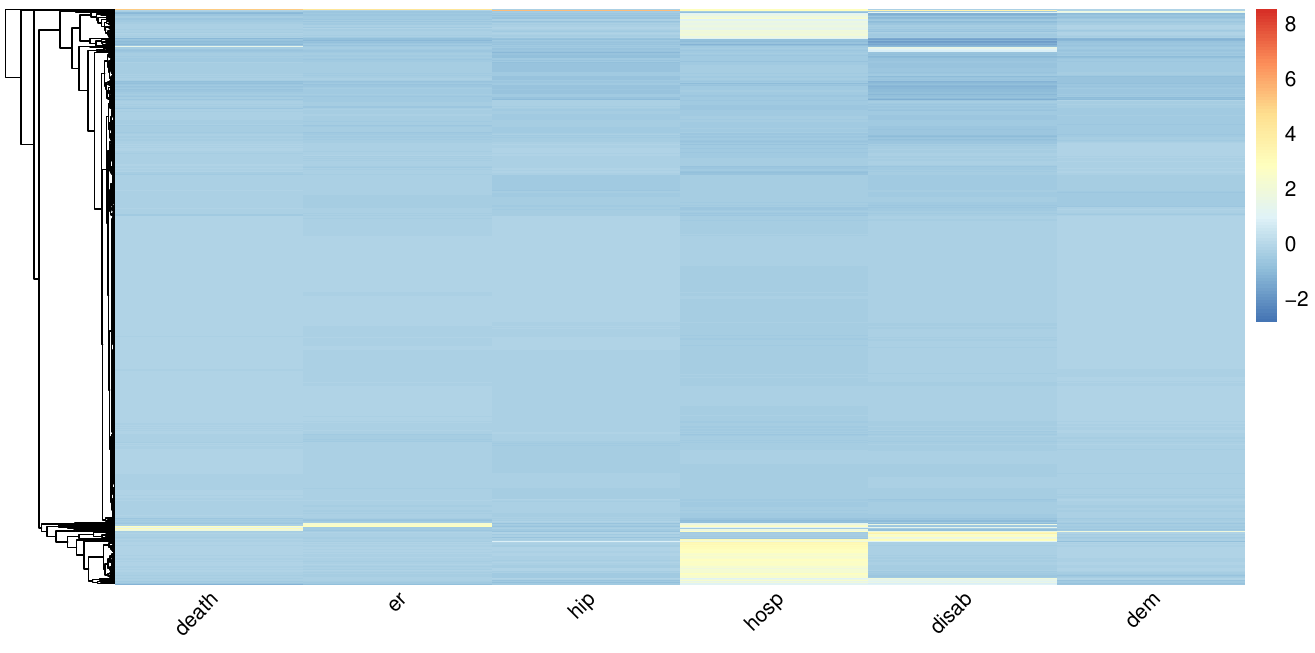}
    \caption{Hierarchical clustering (with Euclidean distance and complete linkage) on the residuals of different logistic models. Acronyms and abbreviations: er: emergency room admission with maximum priority, hip: hip fracture, hosp: hospitalization, disab: disability onset, dem: dementia onset.}
\end{figure}

\clearpage

\section*{S2. On the binomial test for variable selection using the random forest importance}

The variable importance produced by our gradient boosting model can be statistically tested to determine which variables are most significant, similar to what is done in \cite{rachid_zaim_binomialrf_2020} for random forests.
In that framework, a random forest is defined as an ensemble $RF = \{T_1,\dotsi,T_V\}$. Each decision tree $T_v$, with $\space v=1,\dots,V$, is trained on a subsample of both observations and predictors. At each node $k$ of the $v$-th tree a subset of $q < P$ predictors is randomly selected as candidates for splitting, with $q=\sqrt P$. Let $F_{p,v_k}$ be an indicator variable equal to
\begin{equation}
\notag
    F_{p,v_k} = \begin{cases}
        1,\quad \text{if $X_p$ is selected at node $k$ of $T_v$},
        \\
        0, \quad \text{otherwise}.
    \end{cases}
\end{equation}
Thus defined,
$F_{p,v_k} \sim Bern(\theta_p)$, where $\theta_p$ denotes the probability that predictor $X_p$ is selected for splitting at any given node. This probability can be decomposed as the product of (i) the probability that $X_p$ is included in the candidate set at node $k$ ($\pi_{p,v_k}$), and (ii) the conditional probability that $X_p$ is selected as the optimal split among the candidates ($\phi_{p,v_k}$). 

Under the hypothesis that $X_p$ is non-informative, \cite{Ishwaran2010} demonstrated that the probability of being a candidate and selected as a split are independent of the node's depth and position. Therefore (i) and (ii) can be respectively approximated by
$$\pi_{p,v} \approx \frac{q}{P} \qquad \makebox{and} \qquad \phi_{p,v} \approx \frac{1}{q}.$$
Combining (i) and (ii), the authors show that the number of splits $S_p$ for the $p$-th frailty determinant is a binomial random variable $(S_{p,v}|H_0)\sim Binomial(n_v, 1/P)$, where $n_v$ is the number of splits in the $v$-th tree.
To generalize the behaviour at the forest level, the relative frequency of $S_{p,v}$ is computed within each tree and averaged across trees. Under the null hypothesis, this average is expected to be close to $1/P$.

We extended this framework to the gradient tree boosting, where decision trees are weak learners: shallow trees with a limited number of splits \citep{hastie2009elements}. In our setting, we further restrict the model to \textit{decision stumps} (trees with a single split, described for example in \cite{iba1992induction,holte1993very}), ensuring that the probability of selecting a given predictor does not depend on previous splits. Under this specification, a non-informative predictor $X_p$ does not provide a systematic reduction in the loss function across iterations, and its selection can be treated as a random event.
Under these conditions, the selection of a weak variable at any given node remains a primarily a random event.
Therefore, we can maintain the approximation
$$\theta_0 \approx \frac{1}{P}.$$
Let $S_p = \sum_{v=1}^V S_{p,v}$ denote the total count of splits involving $X_p$ across the entire ensemble and let $N_{split} = \sum_{v=1}^V n_v$ be the aggregate number of splits. Under the null hypothesis, $S_p$ can be modelled as $S_p | H_0 \sim Binomial\left(N_{split}, 1/P\right)$.

Therefore, we can test whether $X_p$ is non-informative by comparing the observed splitting frequency of $X_p$ to what would be expected under this null model. Formally, we test
\begin{equation}
    \notag
    H_0: \hat{\theta}_p = \theta_0 \quad \makebox{versus} \quad  H_1: \hat{\theta}_p > \theta_0
\end{equation}
where $\theta_0 = 1/P$ is the baseline probability under the null hypothesis.

\clearpage

\bibliographystyle{abbrvnat} 
\bibliography{reference}